# Accretion of the earliest inner solar system planetesimals beyond the water-snowline

Damanveer S. Grewal[1,2,3,4]*, Nicole X. Nie[1,5], Bidong Zhang[6], Andre Izidoro[7], Paul D. Asimow[1]

[1]Division of Geological and Planetary Sciences, California Institute of Technology, Pasadena, CA 91125, USA
[2]School of Molecular Sciences, Arizona State University, Tempe, AZ 85281, USA
[3]School of Earth and Space Exploration, Arizona State University, Tempe, AZ 85281, USA
[4]Facility for Open Research in a Compressed Environment (FORCE), Arizona State University, Tempe, AZ 85281, USA
[5]Department of Earth, Atmospheric, and Planetary Sciences, Massachusetts Institute of Technology, Cambridge, MA 02139, USA
[6]Department of Earth and Space Sciences, University of California, Los Angeles, CA 90095, USA
[7]Department of Earth, Environmental and Planetary Sciences, Rice University, Houston, TX 77005, USA

*correspondence: damanveer.grewal@asu.edu

**How and where the first generation of inner solar system planetesimals formed remains poorly understood. Potential formation regions are the silicate condensation line and water-snowline of the solar protoplanetary disk. Whether the chemical compositions of these planetesimals align with accretion at the silicate condensation line (water-free and reduced) or water-snowline (water-bearing and oxidized) is, however, unknown. Here we use Fe/Ni and Fe/Co ratios of magmatic iron meteorites to quantify the oxidation states of the earliest planetesimals associated with non-carbonaceous (NC) and carbonaceous (CC) reservoirs, representing the inner and outer solar system, respectively. Our results show that the earliest NC planetesimals contained substantial amounts of oxidized Fe in their mantles (3-19 wt% FeO). In turn, we argue that this required the accretion of water-bearing materials into these NC planetesimals. The presence of substantial quantities of moderately and highly volatile elements in their parent cores is also inconsistent with their accretion at the silicate condensation line and favors instead their formation at or beyond the water-snowline. Similar oxidation states in the early-formed parent bodies of NC iron meteorites and those of NC achondrites and chondrites with diverse accretion ages suggests that the formation of oxidized planetesimals from water-bearing materials was widespread in the early history of the inner solar system.**

Chronological constraints on meteorites, particularly magmatic iron meteorites (sampling the metallic cores of the first generation of planetesimals[1,2]), suggest that planetesimals began forming almost at the onset of solar system formation[3,4]. Planetesimal formation is expected at locations of the disk where small solid particles (mm-to-cm sized pebbles) pile up to sufficient density to collapse under their collective gravity[5,6]. Pebbles are likely to pile up more efficiently in specific regions of the disk – the so-called snowlines – associated with the condensation/sublimation fronts of silicates, water, and carbon monoxide[7–10]. Although there is a growing consensus that planetesimals associated with carbonaceous (CC; sampling the outer solar system) reservoir formed at or beyond the water-snowline[7–10], the formation zone of the first generation of non-carbonaceous (NC; sampling the inner solar system) planetesimals is poorly understood. Understanding the formation of the first NC planetesimals is of utmost importance because Earth and other terrestrial planets chiefly grew from such planetesimals[11,12].

Two competing models have been proposed to explain the formation of the first NC planetesimals. The first is that the formation of NC planetesimals was triggered at the water-snowline during an early phase as the water-snowline migrated out during disk infall (Class I stage), whereas CC planetesimals formed later at the water-snowline as it migrated back in during subsequent evolution of the disk (Class II stage)[7,8]. The alternative view is that NC and CC planetesimals formed contemporaneously at the silicate condensation line and water-snowline, respectively[9,10]. At face value, the latter model is more appealing because magmatic NC irons, based on their chemical compositions, seem to be more reduced than their CC counterparts[4,13,14]. Formation of NC iron meteorite parent bodies (IMPBs) at the silicate condensation line should have resulted in their accreting water-free materials, whereas CC IMPBs forming at the water-snowline should have accreted water-rich materials[9,10]. As water was the primary oxidizing agent in the early forming planetesimals[15–17], accretion of NC and CC IMPBs at the silicate condensation line and water-snowline, respectively, can qualitatively account for their contrasting oxidation states[4]. Accretion of significant amounts of ice by CC IMPBs is also postulated to delay their onset of melting (by lowering the concentration of heat producing $^{26}$Al) and subsequent core formation, relative to the anhydrous NC IMPBs[4]. Therefore, even though the radiometric dating-derived core formation ages of CC IMPBs are ~2 Ma younger than their NC counterparts[1,3], they could have accreted almost contemporaneously[4]. In short, accretion of NC and CC IMPBs at the silicate condensation line and water-snowline, respectively, is an appealing hypothesis given that it appears to be simultaneously consistent with their distinct oxidation states and, potentially, contemporaneous accretion ages. However, whether the differences in chemical characteristics, especially oxidation states, of NC and CC IMPBs are truly consistent with their accretion at the silicate condensation line and water-snowline, respectively, remains elusive because their oxidation states and associated water contents have not been constrained quantitatively. Therefore, whether the first inner solar system planetesimals formed at the silicate condensation line, and thereby accreted reduced and dry materials, needs to be tested.

Here we calculate the oxygen fugacity ($fO_2$) of core-mantle differentiation in magmatic NC and CC IMPBs to constrain the oxidation states of the first generation of planetesimals in each reservoir. Metal-silicate equilibration in partially to fully molten planetesimals established the chemical compositions of the parent cores of magmatic irons[13,18]. At the relatively low pressure and temperatures applicable for core-mantle differentiation in IMPBs, elemental exchange between the metallic cores and silicate mantles is controlled by the $f$O$_2$ of metal-silicate equilibration (M$_{(metal)}$ + n/2 O$_{2(g)}$ = MO$_{n/2(silicate)}$)[18]. Hence, the oxidation states of bulk planetesimals can be determined as long as the compositions of their cores and mantles are known. The bulk compositions of the parent cores of magmatic irons, can be reconstructed by integrating fractional crystallization models and elemental abundances in iron meteorites[13,19–22]. However, the lack of co-existing silicate compositions for magmatic irons presents the principal challenge in constraining the oxidation states of their parent bodies. An alternative approach to constrain the oxidation states of bulk planetesimals during core-mantle differentiation is through the relative depletion of elements in the parent cores that possess similar volatilities but varying degrees of siderophile (metal-loving) character[18]. Fe, Co, and Ni have similar volatilities[23] and hence do not fractionate during condensation/evaporation processes. In contrast, Fe (major element in the cores) can be fractionated relative to Co and Ni at $f$O$_2$ relevant for core-mantle differentiation because of its less siderophile character ($D_{Fe}^{metal/silicate}$ = ~1-10 vs $D_{Ni\ or\ Co}^{metal/silicate}$ = ~100-3000)[24]. This is linked to the positions of the Co-CoO and Ni-NiO buffers, which are ~2 and 4 log units higher in $f$O$_2$, respectively, than the IW buffer (the maximum $f$O$_2$ to form substantial Fe-rich cores in rocky bodies), at temperatures relevant for core-mantle differentiation[18]. The depletion of Fe relative to Ni and Co in the cores can directly constrain the amount of oxidized Fe in the mantles and, as a result, the $f$O$_2$ of the bulk planetesimals relative to the iron-wüstite (IW) buffer.

**Results**

Fractional crystallization models have been previously combined with iron meteorite data to estimate the concentrations of Fe, Ni, and Co in the majority of iron meteorite parent cores (Extended Data Table 1)[13,19–22]. We applied a similar model to calculate the currently unknown $C_{Ni}^{core}$ and $C_{Fe}^{core}$ of group IIIE and $C_{Co}^{core}$ of IC, IIAB, and IIIE cores (Extended Data Fig. 1, Extended Data Table 1; see Methods for details). The (Co/Ni)$_{CI}$ ratios (i.e., the Co/Ni ratio normalized to the ratio in CI chondrites) of the parent cores of both NC and CC irons (except for group IID) are close to 1 (Fig. 1). There are no systematic differences between the (Co/Ni)$_{CI}$ ratios of NC and CC irons. This suggests that there was limited fractionation between Co and Ni during core-mantle differentiation in both NC and CC IMPBs, irrespective of the variation in their oxidation states. The limited fractionation can be readily explained if Co and Ni, in agreement with their high metal-silicate partition coefficients at low pressures[18], almost exclusively partitioned into the cores of both NC and CC IMPBs. This implies near-quantitative reduction of all Co and Ni into the cores without fractionation between them during core-mantle differentiation. (Fe/Ni)$_{CI}$ ratios of the parent cores of both NC and CC irons are, however, lower than 1 (Fig. 1). As reported in previous studies[4,13], CC cores generally have lower (Fe/Ni)$_{CI}$ ratios than NC cores, except for the CC group IIIF, which overlaps the NC groups. Lower Fe contents in the CC cores are a result of a higher retention of Fe as FeO in the mantles of their parent bodies during core-mantle differentiation, implying that CC IMPBs were more oxidized than their NC counterparts[4]. This observation also agrees with the relative enrichment of highly siderophile elements, and consequently, smaller core/mantle mass ratios, in CC IMPBs relative to NC IMPBs[13].

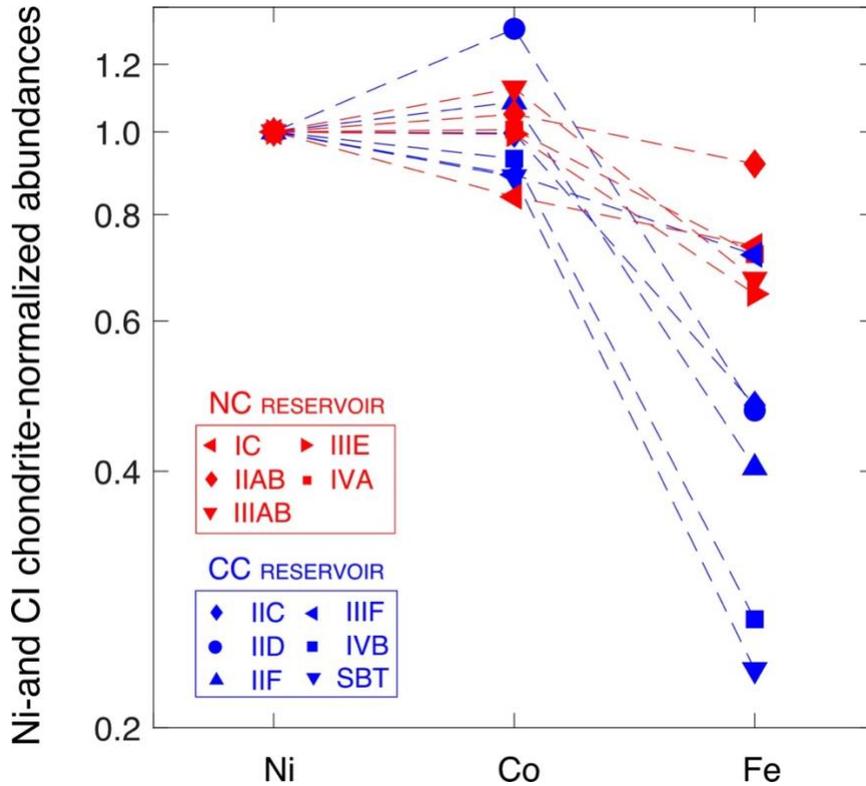

*Figure 1: Ni and CI-chondrite normalized bulk Ni, Co, and Fe contents in the parent cores of magmatic iron meteorites. (Co/Ni)$_{CI}$ ratios of NC and CC irons are close to 1 (except for group IID) and do not show any systematic differences between NC and CC irons. (Fe/Ni)$_{CI}$ ratios of both NC and CC irons are both lower than 1, with CC irons (except for group IIIF) having systematically lower (Fe/Ni)$_{CI}$ ratios than NC irons. NC and CC iron meteorites are shown in red and blue colors, respectively.*

Establishing the oxidation states of NC and CC IMPBs (as $fO_2$ relative to the IW buffer) requires constraints on the amount of oxidized Fe retained in their mantles during core-mantle differentiation. Since the bulk Fe contents of the IMPBs are unknown, a direct mass balance of Fe between mantles and cores cannot be used for this exercise. The bulk Fe/Ni and Fe/Co ratios of chondrites, however, show limited variations across all groups of NC and CC chondrites (standard deviations divided by the means are less than 4%)[25]. As the chondritic Fe/Ni and Fe/Co ratios are also similar to solar values[26], it is reasonable to assume that the earlier forming NC and CC IMPBs also had chondritic Fe/Ni and Fe/Co ratios. Therefore, mass balance of Fe between mantles and cores in tandem with those of Ni and Co (assuming all Ni and Co partition into the cores) can be used to quantify the Fe contents of the mantles of IMPBs by using the equation:

$$C_{Fe}^{mantle} = \left[\left(\frac{Fe}{Ni \text{ or } Co}\right)^{bulk} \times C_{Ni \text{ or } Co}^{core} - C_{Fe}^{core}\right] \times r \qquad \text{(Eq. 1)}$$

where *C* is the concentration of an element in a reservoir and *r* is the core/mantle mass ratio. Core/mantle mass ratios for NC and CC IMPBs were established previously based on the enrichment of highly siderophile elements in their parent cores relative to chondrites[13]. $\left(\frac{Fe}{Ni}\right)^{bulk}$ and $\left(\frac{Fe}{Co}\right)^{bulk}$ ratios of both NC and CC IMPBs were assumed to have CI chondrite-like values (17 and 357, respectively)[25]. With these constraints, the FeO contents of the mantles (assuming all oxidized Fe is FeO) of NC IMPBs computed from Fe/Ni lie in the range 3-16 wt% and those based on Fe/Co are 4-19 wt% (Fig. 2a, Extended Data Fig. 2a, Extended Data Table 2). Likewise, the FeO contents of the mantles of CC IMPBs computed from Fe/Ni are 10-25 wt% and those based on Fe/Co are 6-33 wt% (Fig. 2a, Extended Data Fig. 2a). Hence, the FeO contents of CC IMPBs are generally higher than those of their NC counterparts (except for the overlap of groups IIIF and IVB with NC IMPBs), but the overall differences are small. Lower Fe/Ni and Fe/Co in CC irons are partially offset by their lower core/mantle mass ratios leading to less pronounced differences in the FeO contents of NC and CC IMPBs (Eq. 1). The calculations from Fe/Ni and from Fe/Co are broadly in agreement (Extended Data Fig. 3a), except for groups IC and IID. The source of the discrepancy for these two groups is unknown; it is potentially related to uncertainties in the fractional crystallization models used to estimate the Ni and Co contents of the parent cores or to variations from the CI ratio in the Ni/Co ratios of their parent bodies. Given the limited range in Fe/Ni and Fe/Co ratios across all groups of chondrites[25], our results are not sensitive to the choice of chondrite group used to represent the bulk Fe/Ni and Fe/Co ratios of IMPBs. For example, our results are virtually identical if Fe/Ni and Fe/Co ratios of ordinary chondrites or CM chondrites are assumed for NC and CC IMPBs, respectively, in place of CI chondrites in Eq. 1 (Extended Data Fig. 4).

a)

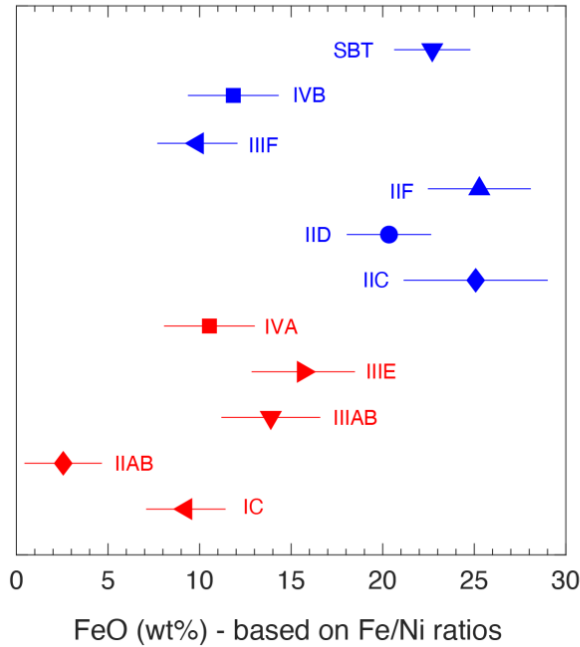

b)

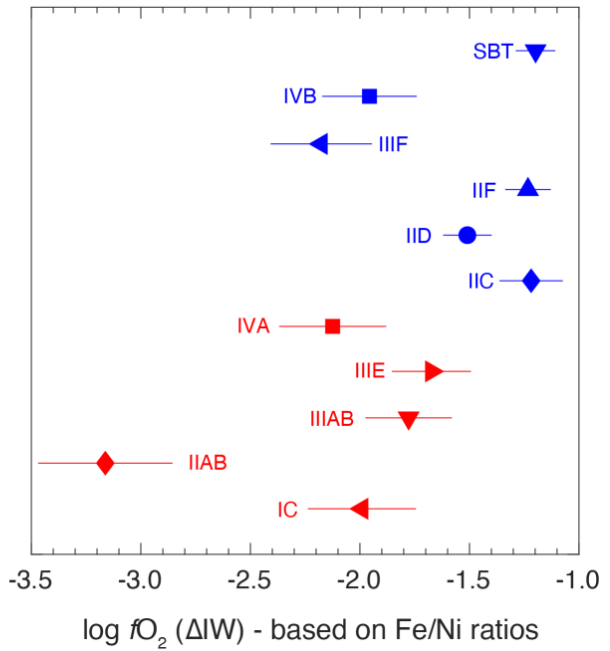

***Figure. 2: Comparison between the FeO contents and fO₂ of IMPBs based on the Fe/Ni ratios of their parent cores.*** *The FeO contents and fO$_2$ of CC IMPBs (blue), except for groups IIIF and IVB, are only modestly higher than those of NC IMPBs (red). Error bars for FeO content and fO$_2$ represent 1σ deviation from the mean obtained by the propagation of standard deviation of individual terms in Eq. 1 and 2, respectively. Uncertainties in the compositions of the parent cores of iron meteorites are not reported in literature[13,19–21]. Therefore, to determine 1σ uncertainties, we assigned a conservative 5% uncertainty each to the concentrations of Fe, Ni and Co in the cores, and bulk Fe/Ni and Fe/Co ratios in CI chondrites. For core/mantle mass ratio, we used the uncertainties reported in ref[13].*

The oxidation states of rocky bodies – represented as *f*O$_2$ relative to the IW buffer (governed by the reaction: Fe$^{metal}$ + ½ O$_2$ = FeO$^{silicate}$) – can be estimated using the following equation:

$$\log fO_2\ (\Delta IW) = 2 \log \frac{a_{FeO}^{silicate}}{a_{Fe}^{metal}} = 2 \log \frac{X_{FeO}^{silicate} \cdot \gamma_{FeO}^{silicate}}{X_{Fe}^{metal} \cdot \gamma_{Fe}^{metal}} \quad (Eq.\ 2)$$

where, $a_{FeO}^{silicate}$, $X_{FeO}^{silicate}$, and $\gamma_{FeO}^{silicate}$ are the activity, mole fraction, and activity coefficient of FeO component in the silicates and $a_{Fe}^{metal}$, $X_{Fe}^{metal}$, and $\gamma_{Fe}^{metal}$ are activity, mole fraction, and activity coefficient of Fe component in the metals[18]. We estimated the *f*O$_2$ values of IMPBs via the ideal solution model ($\gamma_{Fe}^{metal}$ and $\gamma_{FeO}^{silicate}$ = 1) in order to compare our results to the previous reports of *f*O$_2$ in chondrites, achondrites, and rocky planets (Extended Data Table 2). The range of *f*O$_2$ of metal-silicate equilibration in NC IMPBs implied by Fe/Ni is IW−3.2 to IW−1.7, while Fe/Co gives IW−2.8 to IW−1.5 (Fig. 2b, Fig. 3, Extended Data Fig. 2b). By contrast, the range of *f*O$_2$ of metal-silicate equilibration in CC IMPBs based on Fe/Ni is IW−2.2 to IW−1.2 and based on Fe/Co it is IW−2.6 to IW−1.1 (Fig. 2b, Fig. 3, Extended Data Fig. 2b). Although CC IMPBs are modestly more oxidized than NC IMPBs, the *f*O$_2$ of groups IIIF and IVB overlap with their NC counterparts. Note that higher S contents in the parent cores of NC IMPBs[13] result in lower $X_{Fe}^{metal}$, which partially offsets the differences between *f*O$_2$ of NC and CC IMPBs. Just as the mantle FeO contents for each NC and CC IMPB based on Fe/Ni and Fe/Co are similar, the estimates of *f*O$_2$ of metal-silicate equilibration implied from these two ratios are broadly in agreement (variations less than 0.8 log units), except for groups IC and IID (Extended Data Fig. 3b). Although fractional crystallization models yield non-unique solutions to the initial compositions of the parent cores, the calculated FeO content and *f*O$_2$ of each IMPB remain consistent within the range of initial Fe, Ni, and Co contents predicted by previous studies[13,19–22] (Extended Data Fig. 5).

We also extended our findings to a set of ungrouped irons that are presumed to be magmatic (based on their fractionated Ir/Au ratios)[27] and whose heritage from the NC or CC reservoirs has been established previously[28] (Fig. 3; Extended Data Table 3; see Methods for details). Although the bulk compositions of the ungrouped magmatic irons are not as well characterized as those of the grouped magmatic irons, the *f*O$_2$ of core-mantle differentiation in the ungrouped IMPBs from the NC reservoir, is not drastically different from the ungrouped and grouped CC IMPBs (Fig. 3).

Combined, the major conclusion of this exercise is that there are only modest differences between the $fO_2$ prevailing during core-mantle differentiation in NC and CC IMPBs.

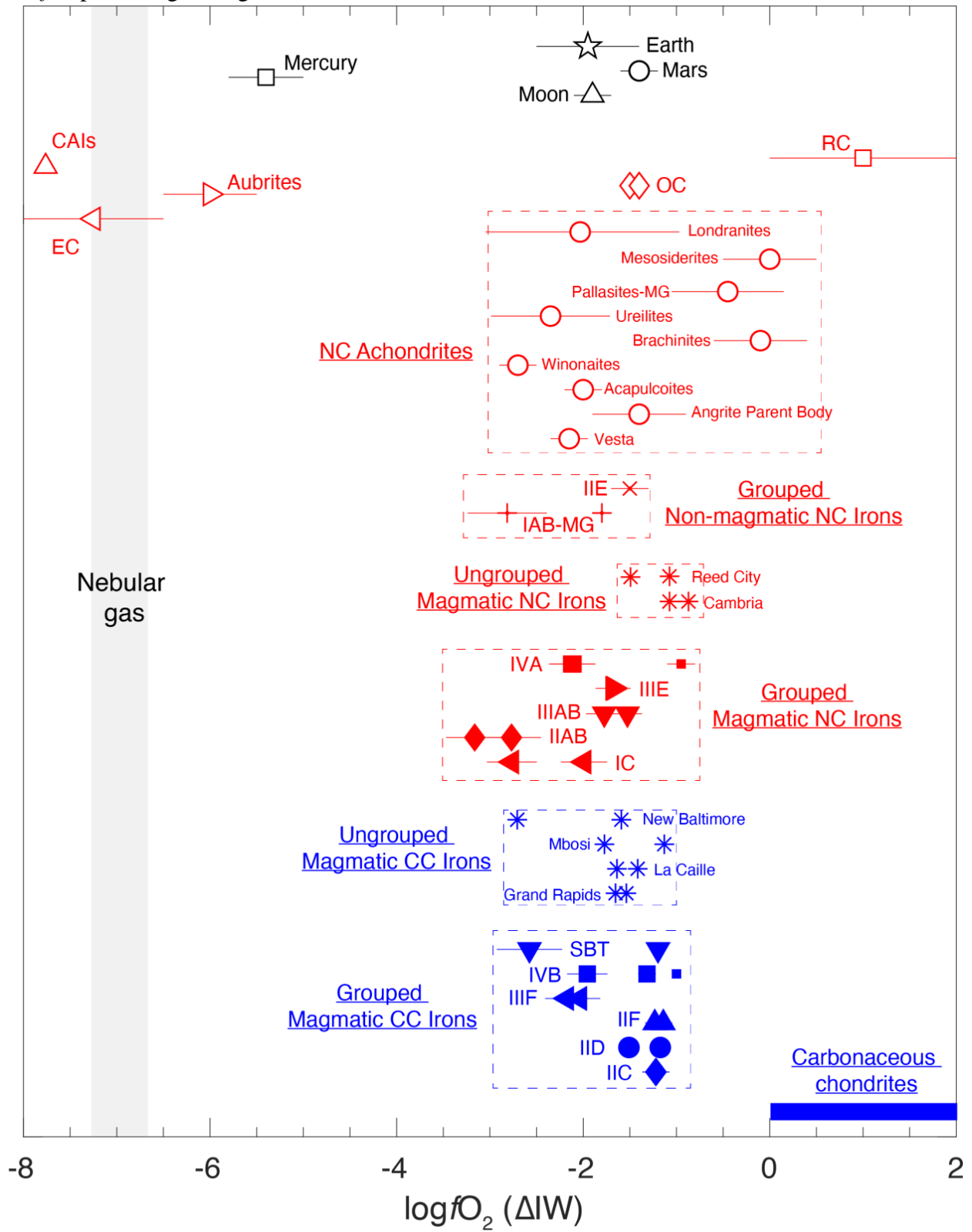

*Figure 3: Variations in the oxidation states of various solar system objects and reservoirs.* The $fO_2$ of all rocky bodies in the NC (red) and CC (blue) reservoirs (except for the enstatite chondrite (EC) and aubrite parent bodies) are several orders of magnitude higher than the nebular gas and CAIs[47]. Grouped NC IMPBs are not drastically more reduced than their CC counterparts. The $fO_2$ of grouped NC IMPBs are similar to those of ungrouped IMPBs, non-magmatic irons, ordinary chondrites (OCs), and several other groups of achondrites from the NC reservoir. The $fO_2$ prevailing during core-mantle differentiation Earth (final event), Mars, and the Moon also lie within the range of grouped NC IMPBs and other oxidized NC planetesimals. The two symbols for each grouped and ungrouped magmatic iron represent $fO_2$ calculated using their Fe/Ni and Fe/Co ratios. The smaller symbols for IVA and IVB groups represent $fO_2$ estimates from previous studies[18,29]. Error bars for grouped and ungrouped magmatic irons represent 1σ deviation from the mean obtained by the propagation of uncertainty from individual terms in the $fO_2$ calculation of Eq. 2. The plotted data are reported in Extended Data Tables 2, 3, and 4.

**Discussion**

*Oxygen fugacity of the NC parent bodies*

Our estimates for $fO_2$ of magmatic NC IMPBs are comparable to previous estimates for some of these bodies (e.g., IW−1.0 for group IVB[18] and IW−0.95 for group IVA[29]) (Fig. 3; Extended Data Table 4). These estimates are also similar to the $fO_2$ of non-magmatic NC iron groups determined by mineral equilibria between the co-existing metals and silicates (e.g., IW−2.8 to IW−1.8 for group IAB-MG[30,31] and IW−1.5 for group IIE[31]). The $fO_2$ of several other primitive and asteroidal NC achondrites are also comparable to our constraints for NC IMPBs. For example, HEDs yield IW−2.2[24,32]; angrites IW−1.4[24,32]; acapulcoites IW−2.0 [30,31]; winonaites IW−2.7[30]; brachinites IW−0.6 to IW+0.4[30,33]; pallasites-MG IW−1.1 to IW+0.9[33]; lodranites IW−3.1 to IW−0.9[33]; ureilites IW−2.4[34]; and mesosiderites IW−0.5 to IW+0.5[35] (Fig. 3; Extended Data Table 4). These values are also comparable to the estimated $fO_2$ of ordinary chondrites (between IW−2.2 and IW−1.5)[31,36] whereas Rumuruti chondrites record an even more oxidized alteration environment (IW to IW+2)[33,37].

All these NC planetesimals, including NC IMPBs, record distinctly higher $fO_2$ than the two groups of extremely reduced NC meteorites, enstatite chondrites (IW−8.0 to IW−6.5[38]) and aubrites (IW−6.5 to IW−5.5[39,40]) (Fig. 3). The Si contents of NC iron meteorites provide further strong evidence that their parent bodies did not accrete from extremely reduced materials akin to enstatite chondrites and aubrites. Si becomes siderophile during metal-silicate equilibration under extremely reduced conditions, such that the cores of planetesimals accreting enstatite chondrite- and aubrite-like materials (containing up to 3 wt% Si in the metal[41,42]) must contain several wt% Si (up to 20 wt%)[39]. Yet Si-bearing iron meteorites are extremely rare; >99% of iron meteorites contain <30 ppm Si in the metal[43,44]. The combined evidence suggests that the oxidation states of NC IMPBs were not drastically different from those of CC IMPBs. The similarity of the oxidation states of NC IMPBs with almost all meteorites from the NC reservoir (except enstatite chondrites and aubrites), covering a wide range of accretion ages (~0.1-4 Ma after CAIs)[4,45] suggests that the formation of oxidized NC planetesimals was the norm rather than the exception.

*Consequences of oxidized NC parent bodies*

High-temperature experiments and thermodynamic calculations predict that equilibrium $fO_2$ during the condensation of nebular solids in the NC reservoir, from a gas with a solar C/O ratio (~0.5), was ~7 log units below the IW buffer[46,47] (Fig. 3). At such reduced conditions, Fe condenses as metallic Fe,Ni-alloy and the co-condensing silicates contain only trace amounts of FeO[46]. $Ti^{3+}/Ti^{4+}$ and $FeO/(FeO + MgO)$ ratios in CAIs attest to equilibrium condensation of nebular solids under such extremely reduced conditions[47] (Fig. 3). By contrast, oxidized Fe is prevalent in the chondrules and matrices of ordinary and Rumuruti chondrites from the NC reservoir. For example, the olivine FeO contents in the type I chondrules of ordinary chondrites can be up to 10 wt% and type II chondrules contain more than 10 wt% FeO[48,49]. These high FeO contents cannot be explained by the evaporation and condensation of anhydrous, FeO-free dust in any known nebular environment[15]. Isotopic compositions of chondrules in ordinary and Rumuruti chondrites also rule out large scale transport of FeO-bearing oxidized dust from the CC to the NC reservoir in the early solar system[50]. Hence, the oxidized nature of early NC planetesimals implies the involvement of an external mechanism to increase $fO_2$ in systems that were otherwise solar in composition.

The first formation of substantial FeO has been linked to environments where the $H_2O/H_2$ ratio was much greater (~$10^{-1}$) than that of nebular gas (~$10^{-4}$)[15,17,51]. Heating of dust in the protosolar nebula with locally enhanced dust/gas ratios is a viable pathway to producing silicates with substantial FeO contents only if the precursor dust was water-bearing – for example, a mixture of solar condensate dust and ice or structurally bound water in phyllosilicates[15,17,52]. An arguably simpler setting is the interaction of aqueous fluids, generated by the melting of ice during $^{26}Al$ decay, with metallic Fe within planetesimals that grew from ice-bearing materials[15,17]. This interaction can form substantial amounts of FeO by the well-characterized reaction: $Fe_{(s)} + H_2O_{(l)} = FeO_{(s)} + H_{2(g)}$[15]. The high porosity and permeability of planetesimals prior to high-temperature sintering allows for the efficient escape of $H_2$ and the ensuing oxidation of the bulk planetesimals[53]. This scenario has also been suggested to explain the production of FeO-bearing chondrule precursors where liquid water interacted with metallic Fe and magnesium silicates at elevated temperatures within planetesimals[15,17].

There is ample evidence supporting a relationship between the presence of abundant oxidized Fe in chondrites (including ordinary and Rumuruti chondrites from the NC reservoir) and a history of aqueous alteration[16,53]. Primitive members of some chondrite groups contain more metal in the matrix and chondrules than their aqueously altered counterparts[53,54]. Oxidation rinds of magnetite[16] surrounding Fe-Ni alloys have also been attributed to parent-body aqueous alteration. In fact, the sequence of increasing $fO_2$ among chondrite classes – from bulk enstatite to ordinary chondrites, Rumuruti chondrites, and carbonaceous chondrites – positively correlates with the water contents and aqueous alteration signatures[55]. It is important to note that, in contrast to chondrites, whose water inventory can potentially be explained by the late addition of water-bearing materials to the surfaces of their parent bodies, the cores of IMPBs record the oxidation state of the entire planetesimals and cannot be a late alteration signature.

We calculated the water contents required to explain the FeO contents of IMPBs via aqueous alteration of anhydrous dust. Production of the entire FeO inventory of NC IMPBs by aqueous alteration requires ~1-5 wt% water equivalent (Extended Data Fig. 6). Note that this estimate is a minimum bound for the water/ice inventory accreted by NC IMPBs because such water can have several other fates aside from oxidizing Fe and being lost as $H_2$ – water may have reacted with other anhydrous components (e.g., matrix)[16] or percolated outwards as aqueous fluid without reacting with the rocky materials[7]. A substantial loss of water inventory during heating/differentiation has also been reported for differentiated CC planetesimals[56]. By comparison, the present $H_2O$/OH contents in ordinary and Rumuruti chondrites are in the range of ~1-4 wt%[55]. However, the primitive water inventories of the parent bodies of ordinary and Rumuruti chondrites are also argued to be much higher than these estimates because of water loss during metamorphic dehydration and consumption of water during oxidation of Fe and other anhydrous components[53,55].

The oxidation states of NC IMPBs, therefore, most likely record a water enriched formation environment via the accretion of either ice or phyllosilicates. Different solar system scenarios can potentially generate such environments. These include:

(1) Formation of NC IMPBs at the water-snowline as it moved inwards, followed by the rapid formation of Jupiter's core[3] beyond the water-snowline separating the NC and CC reservoirs.

(2) The presence of a dynamic water-snowline periodically drifting in and out in the solar protoplanetary disk due to variations in the infall accretion rate onto the sun[57,58]. The transient passage of the water-snowline through the inner part of the disk could have triggered the formation of ice-rich NC IMPBs.

(3) Delivery of oxidized Fe and water via phyllosilicate-bearing dust/pebbles[59] to a pressure bump inside the water-snowline where NC IMPBs could grow. However, the limited thermal stability of phyllosilicates (<800 K)[16] suggests that the silicate condensation line (with a condensation temperature of ~1400 K) cannot be the location of that pressure bump. Pressure bumps associated with poorly understood processes such as long-lived zonal flows[60] could, however, explain the formation of NC IMPBs between the silicate sublimation line and water-snowline where phyllosilicates are stable. However, the efficiency of phyllosilicate formation via nebular processes is currently debated and is generally considered to be a parent body process[16,17], which makes the water-snowline a more plausible location for the formation of NC IMPBs.

*Further implications*

The presence of water-bearing conditions during the accretion of NC IMPBs raises doubts about the predicted contemporaneous accretion ages of NC and CC IMPBs. The ~2 Ma difference in their core-mantle differentiation ages can only be explained if NC and CC IMPBs accreted ice-free and ice-bearing materials, respectively[4]. However, the findings of this study suggest that NC and CC IMPBs did not have drastically different oxidation states and ensuing water contents. These differences are not sufficient to account for the ~2 Ma delay in CC IMPBs' core formation because a substantial difference in water contents (~20-30 wt%) between NC and CC IMPBs would be required for such an explanation[4]. Thus, the observed disparity in their core formation ages likely reflects the later accretion of CC IMPBs[3]. A delayed accretion of CC IMPBs would result in lower $^{26}$Al content, thereby causing lesser dehydration and protracted hydrothermal activity[7], and subsequently greater oxidation of Fe in their bodies.

      A water-bearing character for the building blocks of NC IMPBs does not require that they remained water-rich planetesimals after differentiation. $^{26}$Al decay-powered heating would have resulted in large-scale dehydration of the earliest formed planetesimals[7,61]. As discussed above, $H_2O$ could be lost by outgassing of $H_2$ or by outflow of aqueous fluids. A similar mechanism has been postulated to explain the loss of other highly volatile elements (HVEs) like nitrogen and carbon from these planetesimals[62,63]. Comparably low water contents of NC and CC achondrites also attest to efficient dehydration of planetesimals during differentiation[56]. Consequently, despite the case we have constructed for their elevated initial water contents, the contribution, if any, of these differentiated planetesimals to the water budget of Earth and other terrestrial planets may have been quite limited.

      The presence of nitrogen and carbon in NC iron meteorites suggests that their parent bodies, in addition to water, accreted other HVEs[63–67]. Also, the parent cores of multiple groups of NC IMPBs contained several wt.% S (up to 19 wt%)[13,68] and high amounts of moderately volatile elements (MVEs) like germanium, gallium, arsenic, antimony, and copper, with 50% condensation temperatures lower than 1100 K[2,20]. Had the NC IMPBs accreted at the silicate condensation line, their cores should have been almost free of siderophile HVEs and MVEs, which is not the case. Together with the evidence compiled here for the action of $H_2O$ in producing FeO during or before core-mantle differentiation, the HVE and MVE data argue against formation of early NC planetesimals at the silicate condensation line. Only the enstatite chondrite and aubrite parent bodies plausibly formed in such a hot and reduced water-poor environment. Yet even the enstatite chondrites contain substantial amounts of HVEs and MVEs in both high-temperature and low-temperature components (chondrules and matrix, respectively)[41], which casts doubts on their origin at the silicate condensation line.

      The dichotomy between the oxidation states of the enstatite chondrites and aubrites with those of magmatic irons, non-magmatic irons, ordinary and Rumuruti chondrites, and several other NC achondrites suggests that there were two distinct mechanisms of planetesimal formation in the NC reservoir (Fig. 3). The parent bodies of extremely reduced and water-poor enstatite chondrites and aubrites accreted in the innermost part of the NC reservoir where the water-snowline never reached. All other NC meteorites sampling oxidized planetesimals likely accreted at or beyond the water-snowline. The spread in the accretion ages of oxidized NC meteorites (~0.1-4 Ma after CAIs)[4,45,69] suggests that oxidized planetesimals formed continually in the NC reservoir right from the onset of solar system formation, whereas only late accreting planetesimals (>1.5 Ma after

CAIs; based on the accretion ages of the EC and aubrite parent bodies)[45] are sampled from the reduced part of the NC reservoir. Did enstatite chondrite- and aubrite-like planetesimal formation start late or are early forming planetesimals from that reservoir missing in the meteorite record? Answering this question is critical to understanding the oxidation states and chemical compositions of the seeds of rocky planets like Earth and Mars, which primarily grew from NC planetesimals[11,12,70].

**Methods**
*Fractional crystallization models*

Bulk Fe, Ni, and Co concentrations in the parent cores of groups IIC, IID, IIF, IVB, and SBT are compiled from ref.[19], group IIIF from ref.[22], group IIIAB from ref.[20] and group IVA from ref.[21] (Extended Data Table 1). Bulk Fe and Ni concentrations of groups IC and IIAB cores are compiled from ref[13]. Bulk Co concentrations of IC and IIAB cores are calculated via fractional crystallization models by using the bulk S concentrations of ref.[13] and the lowest Co concentrations (assumed to sample the first crystallized solids) from IC and IIAB groups[71,72]. Bulk S, P, Co, and Ni concentrations are acquired by the fractional crystallization modeling of P-Ni, Ir-Co, and Ir-Ni trends for group IIIE (Extended Data Fig. 1). Ir, Co, and Ni data of IIIE irons are from ref.[73] and P data from ref.[74].

The methods used for fractional crystallization modeling are described in the literature[20,68]. Here is a short summary. The modeling uses batch crystallization in small steps to simulate fractional crystallization of metallic melts. In metallic melts, the partition coefficients of trace elements change as the liquid composition (especially, their S and P concentrations) changes during crystallization. The remaining liquid of a step is used as the starting liquid of the next step.

The equilibrium batch crystallization is a simple mass balance between the phase fields:

$$\frac{C_L}{C_i} = \frac{1}{(1 - f + f \times D_E)} \quad \text{(Eq.3)}$$

In Eq. 3, $C_i$, $C_L$, $f$, and $D_E$ represent the bulk composition of the liquid, the bulk composition of the remaining liquid, the crystallization step, and the partition coefficient of an element (Extended Data Table 5), respectively. The models in this study used a constant of 0.001 for each mass step. The concentration of an element in the solid ($C_s$) that is derived from each mass step is calculated using the bulk composition of the remaining liquid and the partition coefficient of the element in that step:

$$C_s = D_E \times C_L \quad \text{(Eq.4)}$$

The partition coefficient of an element is strongly influenced by the S and P contents of the liquid and varies at each small step. $D_E$ is parameterized using Eq.5:

$$D_E = D_0 \times (Fe\ domains)^\beta \quad \text{(Eq.5)}$$

$D_0$ is the partition coefficient of an element in the S- and P-free system. $\beta$ is a constant specific to an element related to S and P in the liquid. *Fe domains* represent the fraction of free Fe atoms available in the liquid[68]. *Fe domains* in the Fe-Ni-S-P system were calculated by Eq. 6, and $\beta_{S+P}$ of an element in the Fe-Ni-S-P system was calculated using Eq. 7 from ref.[75].

$$Fe\ domains\ = \frac{1 - 2X_S - 4X_P}{1 - X_S - 3X_P} \quad (Eq.6)$$

$$\beta_{S+P} = \left[\frac{2X_S}{(2X_S + 4X_P)}\right]\beta_S + \left[\frac{4X_P}{(2X_S + 4X_P)}\right]\beta_P \quad (Eq.7)$$

$X_S$ and $X_P$ are the molar fractions of S and P in the liquid, respectively. $\beta_S$ and $\beta_P$ are the beta values for each element in the Fe-S and Fe-P systems, respectively.

The scattered interelement trends of group IIIE can be caused by the equilibrium mixing of solid and liquid (trapped melt)[76]. A recently revised version of the trapped-melt model considers the formation of troilite in the trapped melt[20]. The relationship between the trapped melt ($C_{trapped\ melt}$) and the solid ($C_{trapped\ melt\ solid}$) that crystallized from the trapped melt can be expressed using Eq. 8:

$$C_{trapped\ melt\ solid} = \frac{C_{trapped\ melt}}{1 - x} \quad (Eq.8)$$

where $x$ denotes the mass fraction of the trapped melt that solidifies to form troilite.

Ungrouped irons do not belong to any existing iron meteorite groups, and they are either from magmatic cores or non-magmatic impact-melt pools[2,77]. The latter did not go through fractional crystallization. Among the target ungrouped irons, Cambria, Reed City, Grand Rapids, La Caille, Mbosi, and New Baltimore, likely had magmatic origins[77]. We used fractional crystallization models to estimate the bulk compositions of their parent cores. Although these estimates are not as accurate as those for grouped iron meteorites, they can still yield useful points of comparison. Due to the low variation of Ni and Co concentrations within a single core, Ni and Co concentrations of a single iron are approximately equal to those of the first crystallization products of the parent core[68,78]; therefore, we used the Ni and Co concentration of an iron to acquire the bulk composition of its parent core. Fractional crystallization modeling is performed for each ungrouped iron. For NC and CC ungrouped irons, we use 11 wt.% and 3.5 wt% S (mean bulk S content of the parent cores of NC and CC groups, respectively[13]). The acquired bulk Co and Ni concentrations by the modeling are the upper limit of the parent core. Concentrations of Fe are calculated using $C_{Fe} = 100 - C_{Ni} - C_{Co} - C_S$ ("$C$" denotes concentrations of elements).

**Data availability**

The authors declare that the data supporting the findings of this study are available within the article and its Extended Data files.

**Acknowledgements**

Amrita P. Vyas is thanked for helping improve the clarity of our communication. We thank Conel Alexander, Geoff Blake, Konstantin Batygin, and Yoshinori Miyazaki for fruitful discussions during early stages of the research presented in this study. Emily Worsham is thanked for a detailed and constructive review. **Funding:** This study was funded by a Barr Foundation postdoctoral fellowship by Caltech to D.S.G. B.Z. was supported by NASA Grants 80NSSC19K1238 and 80NSSC23K0035.

**Author Contributions**

D.S.G. conceived the project. D.S.G. compiled the data and performed the numerical calculations along with N.X.N. B.Z. performed the fractional crystallization calculations. A.I. helped with the astrophysical implications. All authors interpreted the data. D.S.G. wrote the manuscript with inputs from N.X.N., B.Z., A.I., and P.D.A.

**Competing financial interests**

The authors declare no competing financial and non-financial interests.

**Additional information**

The data that support the plots within this paper and other findings of this study is available in the Extended Data which is available in the online version of the paper. Reprints and permissions information is available at www.nature.com/reprints. Correspondence and requests for materials should be addressed to D.S.G. (damanveer.grewal@asu.edu).

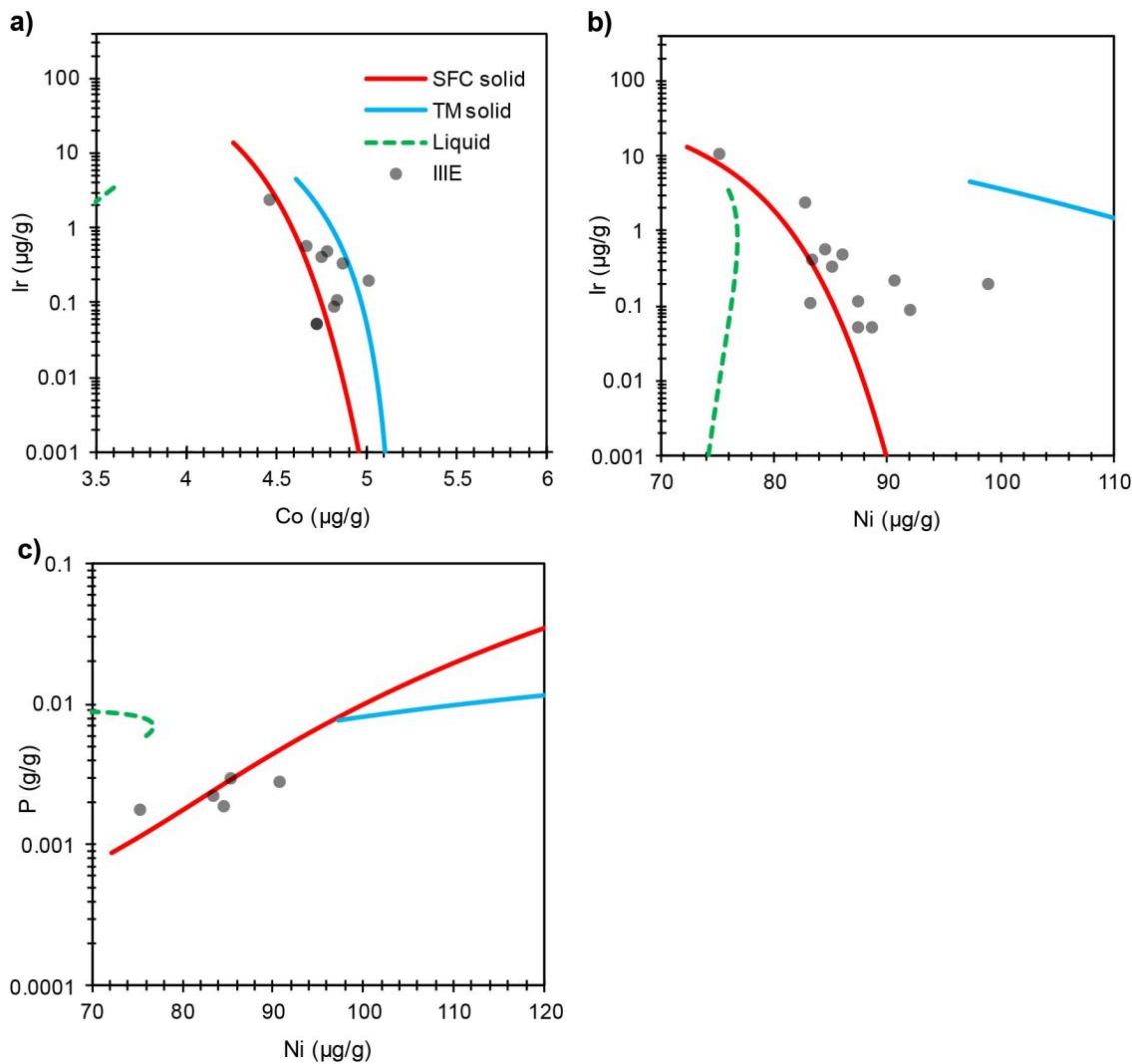

**Extended Data Figure 1**. **Fractional crystallization modeling of P, Ni, Co, and Ir for group IIIE.** The model uses bulk 8 wt.% S and 0.6 wt.% P. The red lines, blue lines, and green dashed lines denote solid from simple fractional crystallization (SFC solid), solid from trapped melt (TM solid), and liquid (Liquid), respectively. Ir, Co, and Ni data are from ref.[73]. Phosphorus data are from ref.[74].

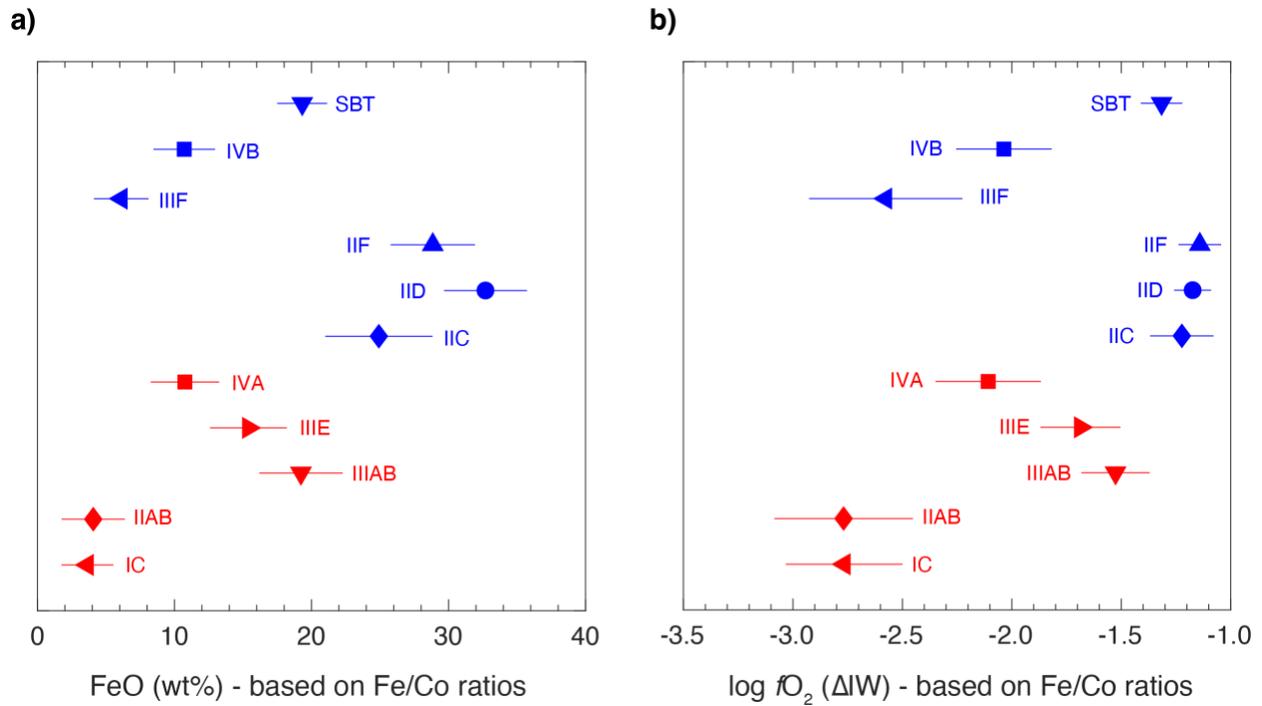

**Extended Data Figure. 2: Comparison between the FeO contents and $fO_2$ of IMPBs based on the Fe/Co ratios of their parent cores.** In agreement with the calculations based on Fe/Ni ratios (Fig. 2), the estimated FeO contents and $fO_2$ of CC IMPBs (blue) based on Fe/Co ratios are either similar to or only modestly higher than those of NC IMPBs (red). Error bars for FeO content and $fO_2$ represent 1σ deviation from the mean obtained by the propagation of standard deviation of individual terms in Eq. 1 and 2, respectively (for details refer to the caption of Fig. 2).

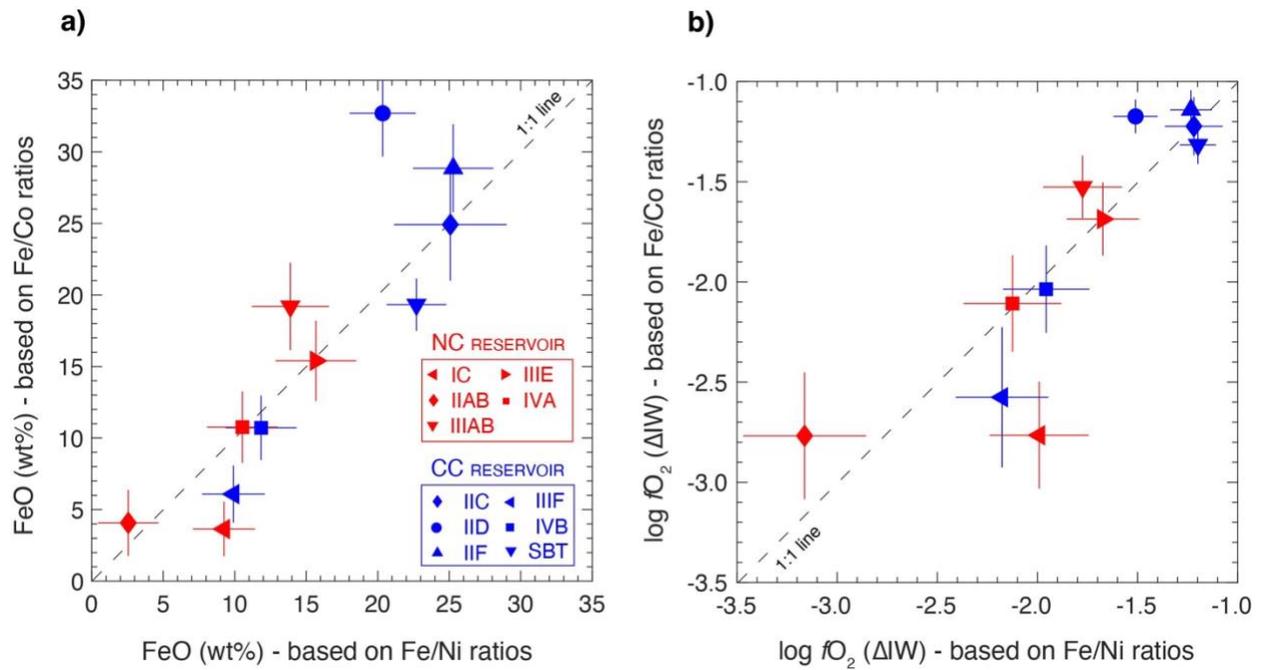

**Extended Data Figure 3: Comparison between the FeO contents and $fO_2$ of IMPBs based on the Fe/Ni and Fe/Co ratios of their parent cores.** The FeO contents and $fO_2$ of each NC and CC IMPB, except for groups IC and IID, estimated via Fe/Ni and Fe/Co ratios of their parent cores broadly agree with each other. Error bars for FeO content and $fO_2$ represent 1σ deviation from the mean obtained by the propagation of standard deviation of individual terms in Eq. 1 and 2, respectively (for details refer to the caption of Fig. 2).

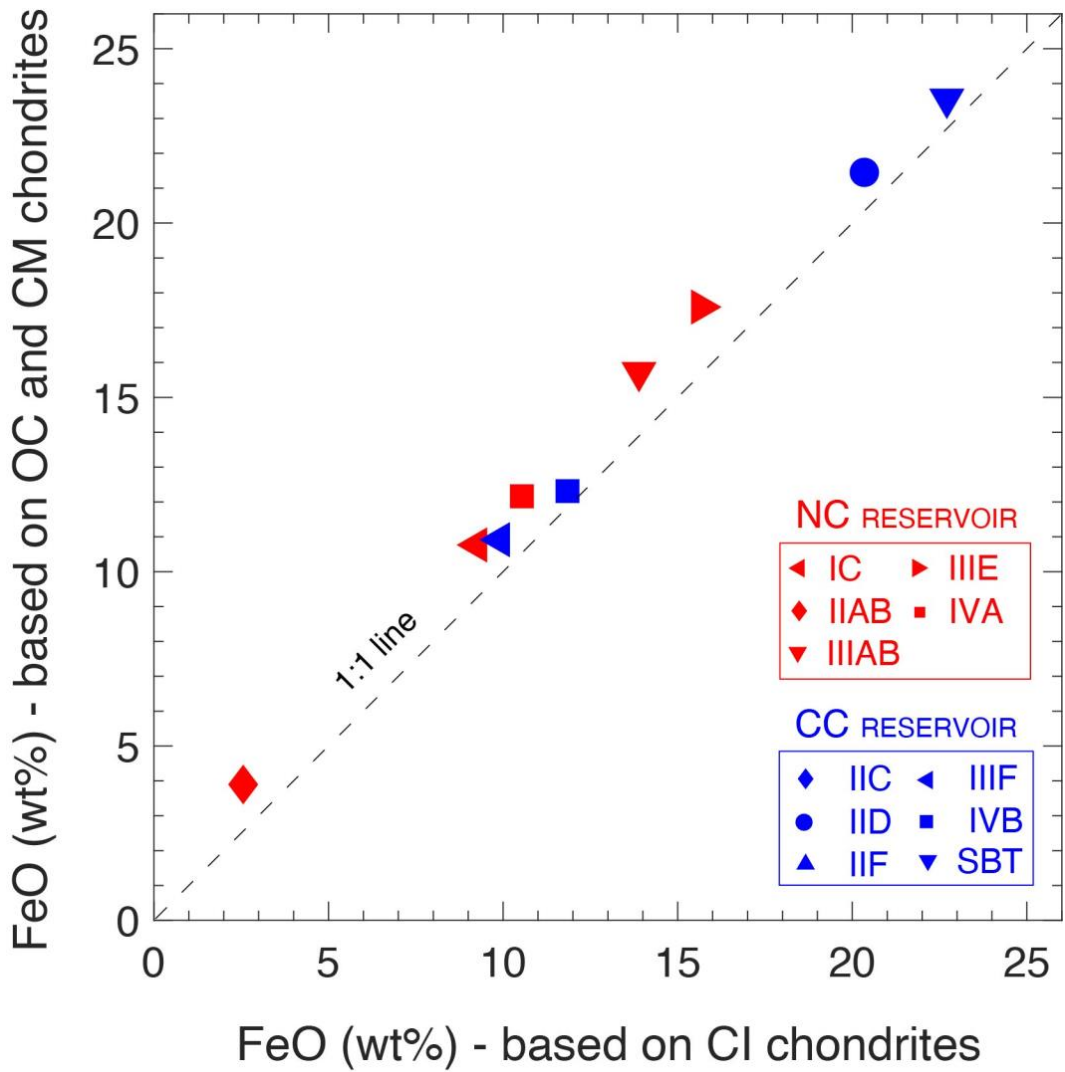

**Extended Data Fig. 4: Comparison between the mean FeO contents of the IMPBs based on the Fe/Ni ratios of CI and ordinary/CM chondrites.** The FeO contents of IMPBs estimated using Fe/Ni ratios of ordinary and CM chondrites (for NC and CC IMPBs, respectively) and CI chondrites are approximately similar.

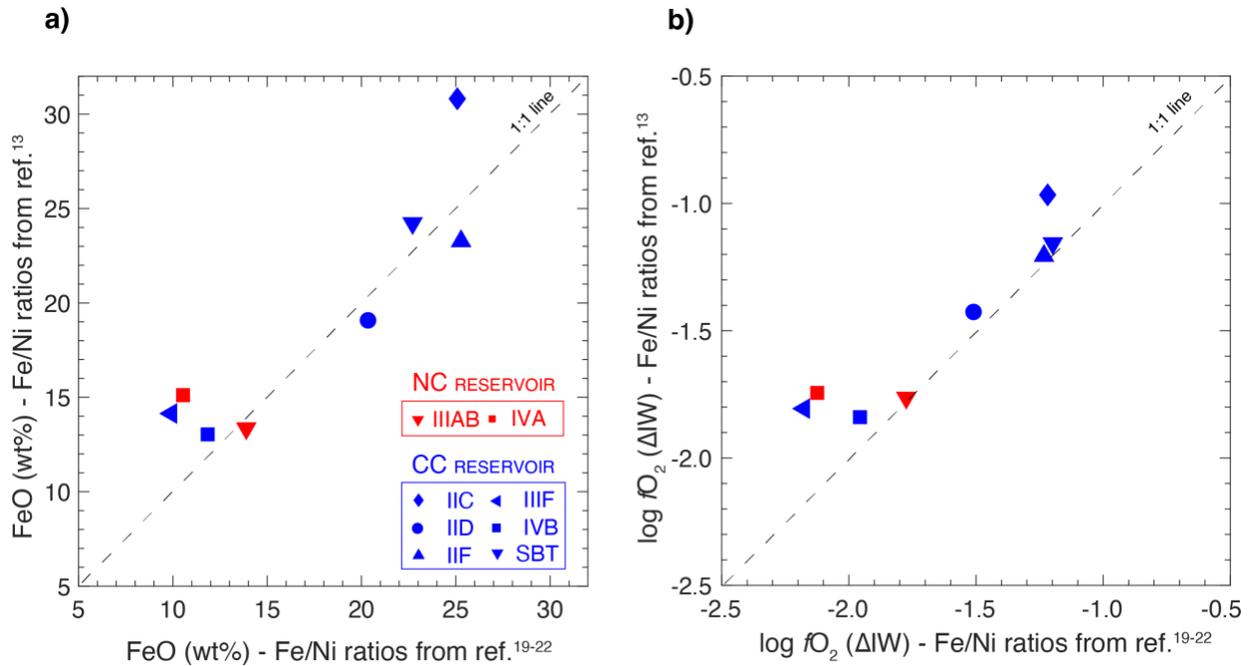

**Extended Data Fig. 5: Comparison between the FeO contents and $fO_2$ of the IMPBs based on the Fe/Ni ratios determined by different fractional crystallization models.** The FeO content and $fO_2$ of each NC and CC IMPB, as determined by several fractional crystallization models, broadly agree with each other. The X-axis represents FeO contents and $fO_2$ values determined using the results from ref.[19–22], while the Y-axis represents values from ref.[13]. Note that the data plotted on the X-axis are used for the discussion in this study. Data for group IC and group IIAB were not plotted since the compositions of their parent cores have only been determined in ref.[13]. Additionally, data for group IIIE was not plotted as the composition of its parent core was determined only in this study.

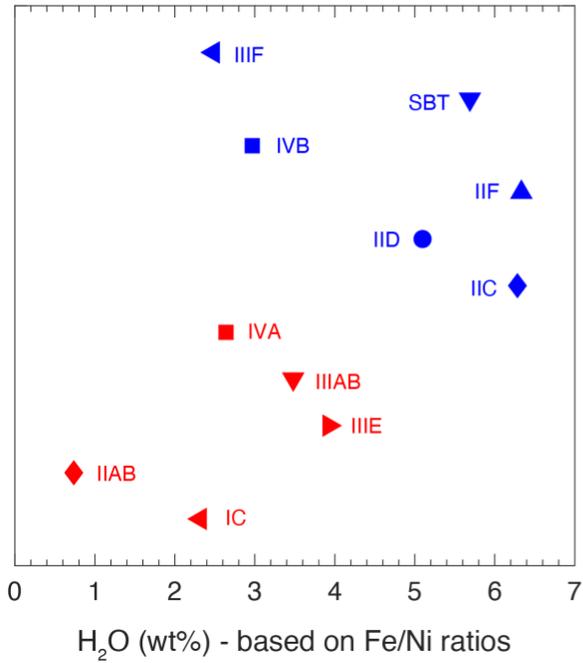 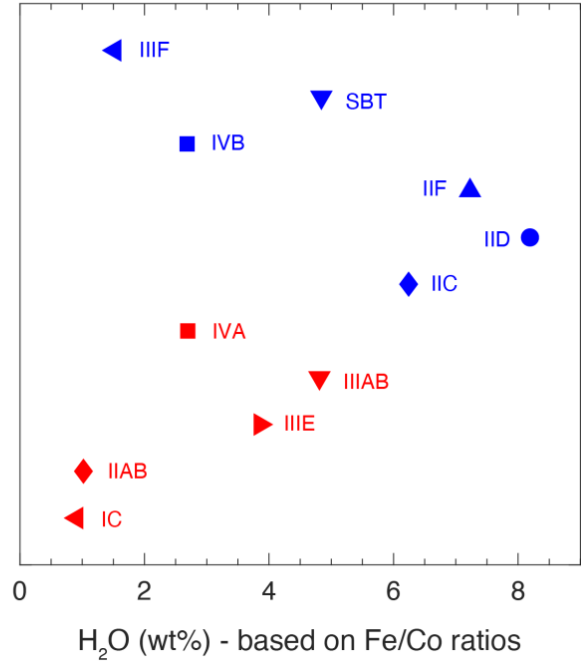

**Extended Data Figure 6: Minimum water required to explain the FeO contents of the parent bodies of iron meteorites based on the Fe/Ni and Fe/Co ratios of their parent cores.** Although CC IMPBs generally require higher water contents than NC IMPBs to explain their FeO contents, the amount of water accreted to explain the FeO contents of NC IMPBs is also substantial.

**Extended Data Table 1: Chemical compositions of the parent cores of NC and CC magmatic iron meteorites.**

| Groups | Fe (wt%) | Ni (wt%) | Co (wt%) | S (wt%) | P (wt%) |
|---|---|---|---|---|---|
| *Magmatic NC Irons* | | | | | |
| IC | 75.0 | 6.0 | 0.2 | 19.0 | 0.1 |
| IIAB | 78.0 | 5.0 | 0.3 | 17.0 | 0.7 |
| IIIAB | 83.0 | 7.3 | 0.4 | 9.0 | 0.2 |
| IIIE | 83.4 | 7.6 | 0.4 | 8.0 | 0.6 |
| IVA | 89.3 | 7.3 | 0.4 | 3.0 | 0.1 |
| *Magmatic CC Irons* | | | | | |
| IIC | 81.3 | 10.0 | 0.5 | 6.0 | 3.0 |
| IID | 86.6 | 10.8 | 0.7 | 0.0 | 1.0 |
| IIF | 81.7 | 11.9 | 0.6 | 5.0 | 0.4 |
| IIIF | 89.1 | 7.3 | 0.3 | 2.0 | 1.0 |
| IVB | 81.0 | 17.8 | 0.8 | 0.0 | 0.7 |
| SBT | 71.7 | 18.0 | 0.8 | 8.0 | 1.0 |

Data sources: Fe, Ni, S, and P contents of groups IC and IIAB – ref.[13]; Co content of groups IC and IIAB – This study; Composition of group IIIAB – ref.[20], IIIE – This study; IVA – ref.[21]; IIC, IID, IIF, IVB, and SBT – ref.[19]; IIIF – ref.[22].

**Extended Data Table 2: FeO contents, $fO_2$, and minimum water contents of the parent bodies of NC and CC magmatic iron meteorites based on Fe/Ni and Fe/Co ratios of their parent cores.**

**Fe/Ni ratios**

| Groups | FeO (wt%) | 1-σ | log$fO_2$ (ΔIW) | 1-σ | $H_2O$ (wt%) |
|---|---|---|---|---|---|
| *Magmatic NC Irons* | | | | | |
| IC | 9.3 | 2.2 | -2.0 | 0.2 | 2.3 |
| IIAB | 2.6 | 2.1 | -3.2 | 0.3 | 0.6 |
| IIIAB | 13.9 | 2.7 | -1.8 | 0.2 | 3.5 |
| IIIE | 15.7 | 2.8 | -1.7 | 0.2 | 3.9 |
| IVA | 10.5 | 2.5 | -2.1 | 0.2 | 2.6 |
| *Magmatic CC Irons* | | | | | |
| IIC | 25.1 | 3.9 | -1.2 | 0.1 | 6.3 |
| IID | 20.3 | 2.3 | -1.5 | 0.1 | 5.1 |
| IIF | 25.3 | 2.8 | -1.2 | 0.1 | 6.3 |
| IIIF | 9.9 | 2.2 | -2.2 | 0.2 | 2.5 |
| IVB | 11.8 | 2.5 | -2.0 | 0.2 | 3.0 |
| SBT | 22.7 | 2.1 | -1.2 | 0.1 | 5.7 |

**Fe/Co ratios**

| Groups | FeO (wt%) | 1-σ | log$fO_2$ (ΔIW) | 1-σ | $H_2O$ (wt%) |
|---|---|---|---|---|---|
| *Magmatic NC Irons* | | | | | |
| IC | 3.6 | 1.9 | -2.8 | 0.3 | 0.9 |
| IIAB | 4.1 | 2.3 | -2.8 | 0.3 | 1.0 |
| IIIAB | 19.2 | 3.0 | -1.5 | 0.2 | 4.8 |
| IIIE | 15.4 | 2.8 | -1.7 | 0.2 | 3.9 |
| IVA | 10.8 | 2.5 | -2.1 | 0.2 | 2.7 |
| *Magmatic CC Irons* | | | | | |
| IIC | 24.9 | 3.9 | -1.2 | 0.1 | 6.2 |
| IID | 32.7 | 3.0 | -1.2 | 0.1 | 8.2 |
| IIF | 28.8 | 3.1 | -1.1 | 0.1 | 7.2 |
| IIIF | 6.1 | 2.0 | -2.6 | 0.3 | 1.5 |
| IVB | 10.7 | 2.2 | -2.0 | 0.2 | 2.7 |
| SBT | 19.3 | 1.8 | -1.3 | 0.1 | 4.8 |

For details related to the calculated values, refer to the main text.

**Extended Data Table 3: Fe, Ni, Co, and S contents of the parent cores, and FeO contents and $fO_2$ of the parent bodies of ungrouped magmatic iron meteorites from the NC and CC reservoirs.**

| Groups | Fe (wt.%) | Ni (wt.%) | Co (wt.%) | S (wt.%) | Fe/Ni | | Fe/Co | |
|---|---|---|---|---|---|---|---|---|
| | | | | | FeO (wt%) | log$fO_2$ ($\Delta$IW) | FeO (wt%) | log$fO_2$ ($\Delta$IW) |
| *Ungrouped Magmatic NC Irons* | | | | | | | | |
| **Cambria** | 78.3 | 10.3 | 0.4 | 11.0 | 32.9 | -0.9 | 25.7 | -1.1 |
| **Reed City** | 80.7 | 7.9 | 0.5 | 11.0 | 18.2 | -1.5 | 31.0 | -1.1 |
| *Ungrouped Magmatic CC Irons* | | | | | | | | |
| **Grand Rapids** | 85.7 | 10.3 | 0.5 | 3.5 | 17.0 | -1.7 | 19.9 | -1.5 |
| **La Caille** | 85.6 | 10.3 | 0.6 | 3.5 | 17.3 | -1.6 | 23.3 | -1.4 |
| **Mbosi** | 86.1 | 9.7 | 0.8 | 3.5 | 15.0 | -1.8 | 36.3 | -1.1 |
| **New Baltimore** | 89.0 | 6.9 | 0.6 | 3.5 | 5.4 | -2.7 | 22.0 | -1.6 |

For details related to the calculated values, refer to the main text and methods section.

**Extended Data Table 4: Oxidation states of rocky bodies in the solar system.**

| Rocky bodies | log$f$O$_2$ ($\Delta$IW) | 1-$\sigma$ | Methodology | Reference |
|---|---|---|---|---|
| Carbonaceous chondrites | 0 to 2 | - | Mineral equilibria | Ref.[33] |
| IVB | -1.0 | - | Assuming a refractory-enriched chondritic silicate composition | Ref.[18] |
| IVA | -1.0 | 0.2 | Cr conc. in irons | Ref.[29] |
| IAB-MG, ung | -2.8 | 0.4 | Quartz-Iron-Ferrosilite reaction | Ref.[30] |
| IAB-MG, ung | -1.8 | - | Coexisting metal, opx and olivine | Ref.[79] |
| IIE | -1.5 | 0.2 | Coexisting metal, opx and olivine | Ref.[79] |
| Vesta | -2.2 | 0.2 | Core-mantle differentiation | Ref.[24] |
| Angrite parent body | -1.4 | 0.5 | Core-mantle differentiation | Ref.[32] |
| Acapulcoites | -2.0 | 0.2 | Coexisting metal, opx and olivine | Ref.[79] |
| Winonaites | -2.7 | 0.2 | Quartz-Iron-Ferrosilite reaction | Ref.[30] |
| Brachinites | -0.1 | 0.5 | V-pre-edge peak in spinels | Ref.[33] |
| Ureilites | -2.4 | 0.6 | Cr valency in olivine cores | Ref.[34] |
| Pallasites | -0.5 | 0.6 | V-pre-edge peak in spinels | Ref.[33] |
| Lodranites | -2.0 | 1.1 | V-pre-edge peak in spinels | Ref.[33] |
| Mesosiderites | 0.0 | 0.5 | Electrochemical Measurements | Ref.[35] |
| H | -1.5 | - | Coexisting metal, opx and olivine | Ref.[79] |
| L | -1.4 | 0.1 | Coexisting metal, opx and olivine | Ref.[79] |
| LL | -1.4 | 0.1 | Coexisting metal, opx and olivine | Ref.[79] |
| RC | 1.0 | 1.0 | Mineral equilibria | Ref.[37] |
| EL3 | -7.3 | 0.8 | Cr valency in olivine | Ref.[38] |
| Aubrite parent body | -6.0 | 0.5 | Basaltic vitrophyres in aubrites | Ref.[39] |
| CAIs | -7.8 | - | Ti valency in fassaite | Ref.[47] |
| Mercury | -5.4 | 0.4 | Core-mantle differentiation | Ref.[80] |
| Moon | -1.9 | 0.2 | Core-mantle differentiation | Ref.[81] |
| Mars | -1.4 | 0.2 | Core-mantle differentiation | Ref.[82] |
| Earth | -2.0 | 0.6 | Final core-mantle differentiation event | Ref.[82] |

Note that the $f$O$_2$ values were reported using an ideal solution model ($\gamma_{Fe}^{metal}$ and $\gamma_{FeO}^{silicate} = 1$).